# Electronegativity Scale from Path Integral Formulation

*Mihai V. Putz*[*] *and Nino Russo*[,]

Dipartimento di Chimica and Centro di Calcolo ad Alte Prestazioni per Elaborazioni Parallele e Distribuite-Centro d'Eccellenza MIUR, Università della Calabria, Via Pietro Bucci, Arcavacata di Rende (CS), I-87030, Italy

[*]*Permanent address:* Chemistry Department, West University of Timisoara, Str. Pestalozzi No.16, Timisoara, RO-1900, Romania.

In the framework of the density functional theory a new electronegativity formulation *via* the Feynman-Kleinert path integral formalism in the markovian limit is proposed. The computation of the electronic density follows, in terms of partition function, the same procedure of the Levy's constrained-search for the wave function. The obtained electronegativity scale seems to respect the main criteria largely used for its acceptability.

**Introduction**

A powerful tool in quantum mechanical and statistical description of many-body systems is given by the path-integral (*PI*) approach. This method, proposed by Feynman in his seminal article [1], provides an alternative formulation of the Schrödinger wave function formalism. The *PI* method is based on the quantum principle of superposition that allows to treat the transition amplitude between two states by the sum of amplitudes along all the possible paths $x(\tau)$ connecting them in a given time $\tau$ [2]. Usually, the paths are parametrized by the imaginary time $i\hbar\beta$, being $\hbar$ the Planck's constant divided by $2\pi$ and $\beta$ the inverse of the thermal energy. When the closed paths are considered, i.e. $x(0) = x(\hbar\beta)$, the *PI* partition function of the considered system in its ground state can be extracted [3]. These are the general features that make the *PI* approach a complete quantum mechanical scheme able to treat both time-dependent and equilibrium properties of a given system in the framework of the so-called Feynman-Kleinert formalism [4-8].

In the last decade the *PI* methods have been the subject of a great interest due to their reliability in treating atoms [9], membranes [10], stochastic [11] and relativistic systems [12], electron-solvent interaction [13-17], polymers [8], electron transfer in transition metal containing systems [18-25] and biological molecules [26-29].

In this work we enlarge the fields of applicability of *PI* methods to the computation of electronegativity in the framework of the density functional theory (*DFT*) [30-33]. It is well known that this concept is crucial in chemistry and has been rigorously defined by Parr et al. in the past [34-38]. The electronegativity, $\chi$, is the primary information of an electronic system and is defined as the ability of atoms to attract electrons to themselves in order to build up molecular systems [39, 40]. In *DFT* the electronegativity is identified as the negative of the chemical potential of a given system [37]. Combining our recently derived analytical expression of the electronegativity [41], with the Feynman-Kleinert *PI* formulation [5], an electronegativity scale for almost the entire periodic system is here given. The obtained results are in good agreement with the previous electronegativity data [35, 36, 42-45], encouraging the future use of the *PI* analytical techniques in the density functional context.

**Theoretical Method**

*Density Functional Electronegativity Formulation*

In *DFT* the electronegativity assumes the general form [37]:

$$\chi(N) = -\left(\frac{\partial E}{\partial N}\right)_V \qquad (1)$$

being *E* and *N* the total energy and number of electrons at constant external potential *V*. The above so called absolute electronegativity recovers the chemical Mulliken electronegativity, $\chi_M$ [40], through out the identity [46]:

$$\chi_M = \frac{1}{2}\int_{N-1}^{N+1} \chi(N) dN . \qquad (2)$$

Nevertheless from equations 1 and 2 appears that to compute the Mulliken electronegativity the total energy expression is required. Within the wave-function approach of electronic density, since the ground-state energy has to be found, the specific Levy constrained-search scheme is applied [47]. This procedure involves the minimization of the trial total energy in two steps: by the first one all the possible wave functions are included into the classes that give the same density, while the second step consists in a overall minimization of the densities. This scheme is of fundamental importance in *DFT* with the merit to transform the (unsolved) *V*-representability problem into the *N*-representability one [32]:

$$\rho(x) \geq 0$$

$$\int \left|\nabla \rho(x)^{1/2}\right|^2 dx < \infty$$

$$\int \rho(x) dx = N . \qquad (3)$$

Recently, [41], employing the softness kernel version of *DFT* [48], it was introduced an analytical expression of the Mulliken electronegativity by-passing the total energy expression. In atomic units it looks like (see Appendix):

$$\chi_M = \frac{b+N-1}{2\sqrt{a}}\arctan\left(\frac{N-1}{\sqrt{a}}\right) - \frac{b+N+1}{2\sqrt{a}}\arctan\left(\frac{N+1}{\sqrt{a}}\right) + \frac{C_A - 1}{4}\ln\left[\frac{a+(N-1)^2}{a+(N+1)^2}\right] \quad (4)$$

where in order to simplify the expression, the following definitions have been introduced:

$$a = \int_{-\infty}^{+\infty} \frac{\nabla\rho(x)}{[-\nabla V(x)]} dx$$

$$b = \int_{-\infty}^{+\infty} \frac{\nabla\rho(x)}{[-\nabla V(x)]} V(x) dx$$

$$C_A = \int_{-\infty}^{+\infty} \rho(x) V(x) dx . \quad (5)$$

The first two quantities depend on the density gradient and can be termed as *chemical response* indices, whereas $C_A$, which is the simple convolution between the density and the external potential represents the *chemical action* index [41]. Moreover, the electronic density in equation 4 has to fit with the Levy's algorithm, i.e. to be *N*-representable.

The electronegativity expression given in the equation 4 represents our proposed *DFT* electronegativity working formula.

*Path Integral Constrained-Search Algorithm*

The Mulliken electronegativity requires the knowledge of the electronic density under the external potential influence. The computation of the electronic density can be carried out using the following expressions:

$$\rho_1(x_0) = Z_1^{-1} \frac{1}{\sqrt{2\pi\hbar^2\beta/m_0}} \exp[-\beta W_1(x_0)] . \quad (6)$$

$$Z_1 = \frac{1}{\sqrt{2\pi\hbar^2\beta/m_0}} \int_{-\infty}^{+\infty} dx_0 \exp[-\beta W_1(x_0)] \qquad (7)$$

where the path influence is comprised within the introduced Feynman centroid $x_0$ [49-51].

Since the Feynman-Kleinert *PI* approach is based on the optimization of the variational condition [5]:

$$Z \geq Z_1 \qquad (8)$$

the partition function $Z_1$, related to the approximate potential $W_1(x_0)$, has to be as close as possible to the effective partition function Z of the applied external potential $V(x_0)$. The condition 8 can be seen as the counterpart of the first step in Levy's algorithm in terms of partition function: among the possible effective partition functions is chosen that one that closely approximates the potential $W_1(x_0)$ to the real Hamiltonian in equation 7. Following this variational procedure an optimal form is found for the potential $W_1(x_0)$ [5-8]:

$$W_1(x_0) = \frac{1}{\beta}\log[\frac{\sinh(\hbar\beta\Omega(x_0)/2)}{\hbar\beta\Omega(x_0)/2}] + V_{a^2(x_0)}(x_0) - \frac{m_0}{2}\Omega^2(x_0)a^2(x_0) \qquad (9)$$

in which the components $V_{a^2(x_0)}(x_0)$, $\Omega^2(x_0)$ and $a^2(x_0)$ are expressed as:

$$V_{a^2(x_0)}(x_0) = \int_{-\infty}^{+\infty}\frac{dx'_0}{\sqrt{2\pi a^2(x_0)}}V(x'_0)\exp[-\frac{(x'_0-x_0)^2}{2a^2(x_0)}] \qquad (10)$$

$$a^2(x_0) = \frac{1}{m_0\beta\Omega^2(x_0)}[\frac{\hbar\beta\Omega(x_0)}{2}\coth(\frac{\hbar\beta\Omega(x_0)}{2}) - 1] \qquad (11)$$

$$\Omega^2(x_0) = \frac{2}{m_0}\frac{\partial V_{a^2(x_0)}(x_0)}{\partial a^2(x_0)}. \qquad (12)$$

It is worth nothing that equations 11 and 12 arise through out the minimization of the equation 9 with respect to the parameters $\Omega(x_0)$ and $a^2(x_0)$. As a consequence, also the second step of the Levy's constrained-search algorithm is achieved: among the possible potentials $W_1$ is found out the potential which closely approximate the entire Hamiltonian of the system. So, the Feynman-Kleinert *PI* method provides a suitable analytical counterpart, in terms of partition function, to the Levy's formalism in terms of wave function.

The PI procedure will be applied in the framework of the so-called Markovian approximation [50] ($\beta \to 0$) which will cancels the low temperature quantum fluctuations. This limit corresponds also to the ultra-short correlation of the involved electrons with the applied external potential due to the time dependence of the quantum statistical quantity $\hbar\beta \propto \Delta t$. This means that assuming the free motion of the electrons in absence of an external potential ($\Delta t = 0 \Leftrightarrow \beta = 0$), as the external potential is applied an immediate orbit stabilization of the electronic system is reached ($\Delta t \to 0 \Leftrightarrow \beta \to 0$). In other words, the escape (unstable) paths are precluded [52, 53]. Finally, this limit introduces also *correlation effects* with the medium. Changing the variable in such a way that $z(x'_0) = (x'_0 - x_0)/\sqrt{2a^2(x_0)}$ and $dz = dx'_0/\sqrt{2a^2(x_0)}$ in equation 10, the smeared potential can be written in terms of the so called Wigner expansion [54] of a high temperature limit ($\beta \to 0$):

$$\begin{aligned} V_{a^2}(x_0) &= \frac{1}{\sqrt{2\pi a^2(x_0)}} \int_{-\infty}^{+\infty} V(x'_0) \exp\left[-\frac{(x'_0 - x_0)^2}{2a^2(x_0)}\right] dx'_0 \\ &= \frac{1}{\sqrt{2\pi a^2(x_0)}} \sqrt{2a^2(x_0)} \int_{-\infty}^{+\infty} V\left(x_0 + \sqrt{2a^2(x_0)}z\right) \exp(-z^2) dz \\ &= \frac{1}{\sqrt{\pi}} \int_{-\infty}^{+\infty} \left\{V(x_0) + \sqrt{2a^2(x_0)}zV'(x_0) + \frac{1}{2}(2a^2(x_0))z^2 V''(x_0) + ..\right\} \exp(-z^2) dz \\ &\cong \frac{1}{\sqrt{\pi}} \int_{-\infty}^{+\infty} \left\{V(x_0) + \frac{1}{2}(2a^2(x_0))z^2 V''(x_0)\right\} \exp[-z^2] dz \\ &= V(x_0) + \frac{1}{2}a^2(x_0) V''(x_0) \end{aligned} \qquad (13)$$

Now, the optimized frequency given in equation 12 in this limit becomes:

$$\Omega^2(x_0) \cong \frac{1}{m_0} V''(x_0) . \tag{14}$$

and the equation 11 is simplified as:

$$a^2(x_0) \cong \hbar^2 \frac{\beta}{12 m_0} . \tag{15}$$

From equations 6, 7, 9 and 13-15 appears that the Feynman-Kleinert *PI* constrained-search algorithm in the markovian limit provides an efficient recipe to compute the electronic densities using only the external potential dependence.

Parr and Yang have shown [32] that the integral formulation of the Kohn-Sham *DFT* arrives to the electronic density expression performing Wigner semi-classical expansion combined with the short time approximation (in $\beta$ parameter). A similar relation arises from our approach in which effective potential is approximated through the equation 9. All the components around *V(x)* can be formally interpreted as the *exchange-correlation PI potential* $V_{XC}^{PI}(x)$ with the medium. Even if this potential can be expanded in higher orders, the truncation of the expansion to the second order [38] gives:

$$W_1(x_0)_{\beta \to 0} \cong V(x_0) + \hbar^2 \frac{\beta}{24 m_0} V''(x_0)$$

$$\equiv V(x_0) + V_{XC}^{PI}(x_0) \tag{16}$$

in which the exchange-correlation *PI* potential with the medium:

$$V_{XC}^{PI}(x_0) = \hbar^2 \frac{\beta}{24 m_0} V''(x_0) \tag{17}$$

corrects the classical external potential *V(x)*.

Finally it is worth to note that instead of equation 3, from equations 6 and 7 the normalization condition looks like:

$$\int_{-\infty}^{+\infty} \rho_1(x_0) dx_0 = 1 . \tag{18}$$

**Computational Details**

In order to apply the proposed Mulliken electronegativity formula (see equation 4), the core potential, in which the valence electrons are moving should be known. This information can be obtained from the pseudopotential approach. In particular, in our work, the Stuttgart/Bonn pseudo-potentials have been employed [55].

Since the electronic density depends on the $\beta$ parameter, in the *PI* calculations it can be fixed in such a manner that the electronic density fulfills the normalization condition 18. The computations have been performed in the markovian limit, $\beta \to 0$, and a scale factor has been introduced for the electronegativity and the chemical action in order to have similar results (in absolute values). In this way the relationship between the chemical action and electronegativity recovers the previous one [56, 57]:

$$\chi(N,Z) = \left\langle \frac{1}{x} \right\rangle = \int \left\{ \rho(N,Z,x) \frac{1}{x} \right\} dx = -\int \left\{ \rho(N,Z,x) V_{COULOMB}(x) \right\} dx \equiv -C_A^{COULOMB} \quad (19)$$

with *Z* equal to the nuclear charge.

**Results and Discussion**

Our results are reported in Table 1 for chemical action and in Table 2 for electronegativity together with some previous electronegativity scales obtained theoretically [35, 42, 43] or from experimental ionisation potentials and electron affinities [36, 44].

The proposed electronegativity scale follows the general rules for its acceptability [58]. The decreasing of $\chi$ along the group is respected (see, for instance, Ga<Al and Ge<Si) as well as its difference in going from light to heavy atoms of the same group. $\chi$ increases left to right across rows taking into account that for some heavy elements the relativistic effects, which are not considered in the computations, can affect this trend. Correctly, the halogen atoms have the highest electronegativity with respect to their left row neighbours. Looking to the transition metal atoms, we underline that the obtained electronegativities fall in a narrow range of values compared with those of the main group atoms.

**Table 1.** The atomic chemical action values (in eV) computed by path integral method.

| Li | Be | | | | | | | | | | | B | C | N | O | F | Ne |
|---|---|---|---|---|---|---|---|---|---|---|---|---|---|---|---|---|---|
| 4.77 | 6.05 | | | | | | | | | | | 6.77 | 8.69 | 9.73 | 10.93 | 11.84 | 10.90 |
| Na | Mg | | | | | | | | | | | Al | Si | P | S | Cl | Ar |
| 4.09 | 5.18 | | | | | | | | | | | 8.73 | 5.95 | 8.38 | 9.48 | 9.94 | 9.25 |
| K | Ca | Sc | Ti | V | Cr | Mn | Fe | Co | Ni | Cu | Zn | Ga | Ge | As | Se | Br | Kr |
| 3.28 | 4.41 | 2.66 | 3.19 | 3.78 | 4.71 | 5.41 | 5.35 | 5.39 | 5.49 | 5.83 | 4.54 | 3.24 | 5.12 | 4.53 | 9.09 | 9.11 | 7.93 |
| Rb | Sr | Y | Zr | Nb | Mo | Tc | Ru | Rh | Pd | Ag | Cd | In | Sn | Sb | Te | I | Xe |
| 1.63 | 2.92 | 3.04 | 3.57 | 4.34 | 5.08 | 5.06 | 5.36 | 5.65 | 5.86 | 5.86 | 4.76 | 5.10 | 5.37 | 5.05 | 7.53 | 8.42 | 7.37 |

The six considered metalloid elements (B, Si, Ge, As, Sb, Te), that separate the metals from the non-metals, have electronegativity values, which do not allow overlaps between metals and non-metals. Furthermore, looking at the $\chi$ metal values the requirement that they must have electronegativities lower than silicon is satisfied (see for example Ga<Si, Al<Si, Ge<Si) following the so-called silicon rule [58]. Finally, we briefly discuss the values obtained for the N, O, F, Ne, Cl, Ar, Br, and Kr elements that present oxidation states lower than their valence electrons. The rule in this case states that $\chi$ parallels the decreasing in valence electrons. The results follow this rule with the exception of the chlorine atom that has a $\chi$ value higher than the nearest noble gas atom Ar. The electronegativity trend for these atoms results in the order Ne>F>O>Cl>N>Ar>Kr>Br.

**Table 2.** The atomic Mulliken electronegativities (in eV) computed by path integral method and from different levels of theory and experiment.

| Element | Mulliken-Jaffe[a] | Experiment[b] | Xα[c] | Present Work |
|---|---|---|---|---|
| Li | 1.8 | 3.01 | 2.58 | 4.11 |
| Be | 4.8 | 4.9 | 3.80 | 5.64 |
| B | 5.99 | 4.29 | 3.40 | 5.72 |
| C | 8.59 | 6.27 | 5.13 | 8.56 |
| N | 11.21 | 7.27 | 6.97 | 10.13 |
| O | 14.39 | 7.53 | 8.92 | 11.87 |
| F | 12.18 | 10.41 | 11.0 | 13.13 |
| Ne | 13.29 | - | 10.31 | 13.39 |
| Na | 1.6 | 2.85 | 2.32 | 3.16 |
| Mg | 4.09 | 3.75 | 3.04 | 4.52 |
| Al | 5.47 | 3.21 | 2.25 | 5.80 |
| Si | 7.30 | 4.76 | 3.60 | 6.56 |
| P | 8.90 | 5.62 | 5.01 | 9.04 |
| S | 10.14 | 6.22 | 6.52 | 10.09 |
| Cl | 9.38 | 8.30 | 8.11 | 10.64 |
| Ar | 9.87 | - | 7.11 | 10.12 |
| K | 2.90 | 2.42 | 1.92 | 3.15 |
| Ca | 3.30 | 2.2 | 1.86 | 4.21 |
| Sc | 4.66 | 3.34 | 2.52 | 2.93 |
| Ti | 5.2 | 3.45 | 3.05 | 3.52 |
| V | 5.47 | 3.6 | 3.33 | 4.19 |
| Cr | 5.56 | 3.72 | 3.45 | 5.23 |
| Mn | 5.23 | 3.72 | 4.33 | 6.02 |
| Fe | 6.06 | 4.06 | 4.71 | 5.96 |
| Co | 6.21 | 4.3 | 3.76 | 6.01 |
| Ni | 6.30 | 4.40 | 3.86 | 6.12 |
| Cu | 6.27 | 4.48 | 3.95 | 6.35 |
| Zn | 5.53 | 4.45 | 3.66 | 5.07 |
| Ga | 6.02 | 3.2 | 2.11 | 3.49 |
| Ge | 6.4 | 4.6 | 3.37 | 5.45 |
| As | 6.63 | 5.3 | 4.63 | 4.87 |
| Se | 7.39 | 5.89 | 5.91 | 7.71 |
| Br | 8.40 | 7.59 | 7.24 | 7.75 |
| Kr | 8.86 | - | 6.18 | 8.65 |
| Rb | 2.09 | 2.34 | 1.79 | 1.56 |
| Sr | 3.14 | 2.0 | 1.75 | 2.87 |
| Y | 4.25 | 3.19 | 2.25 | 3.33 |
| Zr | 4.57 | 3.64 | 3.01 | 3.92 |
| Nb | 5.38 | 4.0 | 3.26 | 4.77 |
| Mo | 7.04 | 3.9 | 3.34 | 5.59 |
| Tc | 6.27 | - | 4.58 | 5.57 |
| Ru | 7.16 | 4.5 | 3.45 | 5.91 |
| Rh | 7.4 | 4.3 | 3.49 | 6.23 |
| Pd | 7.16 | 4.45 | 3.52 | 6.46 |
| Ag | 6.36 | 4.44 | 3.55 | 6.47 |
| Cd | 5.64 | 4.43 | 3.35 | 5.26 |
| In | 5.22 | 3.1 | 2.09 | 5.38 |
| Sn | 6.96 | 4.30 | 3.20 | 5.75 |
| Sb | 7.36 | 4.85 | - | 5.44 |
| Te | 7.67 | 5.49 | 5.35 | 6.35 |
| I | 8.10 | 6.76 | 6.45 | 7.12 |
| Xe | 7.76 | - | 5.36 | 7.80 |

[a] Ref. [42, 43]; [b] Ref. [36, 44]; [c] Ref. [35]

It is worth to note that the *PI* treatment does not need the orbital type function but only the pseudo-potential representing the field in which the electrons move. In order to verify the influence of the different orbital type we have redone the electronegativity computation for C, O and N atoms by using p-type orbitals and the sp, $sp^2$ and $sp^3$ hybridisation states. Results, reported in Table 3 and Figure 1, show how the actual electronegativity formulation preserves also the orbital hierarchy and is sensitive to the hybrid orbitals as well. Finally, analysing Table 3 and Fig.1 we underline that the electronegativity trend from a type of hybridisation to another is similar.

**Table 3.** The orbital electronegativities (in eV) and the absolute chemical actions (in eV) for C, N and O atoms.

|   |   | s | p | sp | $sp^2$ | $sp^3$ |
|---|---|---|---|---|---|---|
| C | $\chi^{MJ\ a}$ | 8.59 | 5.80 | 10.39 | 8.79 | 7.98 |
|   | $\chi$ | 8.56 | 4.04 | 9.89 | 6.99 | 5.71 |
|   | $-C_A$ | 8.69 | 4.1 | 10.04 | 7.1 | 5.71 |
| N | $\chi^{MJ\ a}$ | 11.21 | 7.39 | 15.68 | 12.87 | 11.54 |
|   | $\chi$ | 10.13 | 6.14 | 17.54 | 12.40 | 10.13 |
|   | $-C_A$ | 9.73 | 5.9 | 16.86 | 11.92 | 9.73 |
| O | $\chi^{MJ\ a}$ | 14.39 | 9.65 | 27.25 | 17.07 | 15.25 |
|   | $\chi$ | 11.87 | 8.39 | 27.40 | 19.38 | 15.82 |
|   | $-C_A$ | 10.93 | 7.73 | 25.23 | 17.84 | 14.57 |

[a] Mulliken-Jaffè electronegativity, Ref. [45]

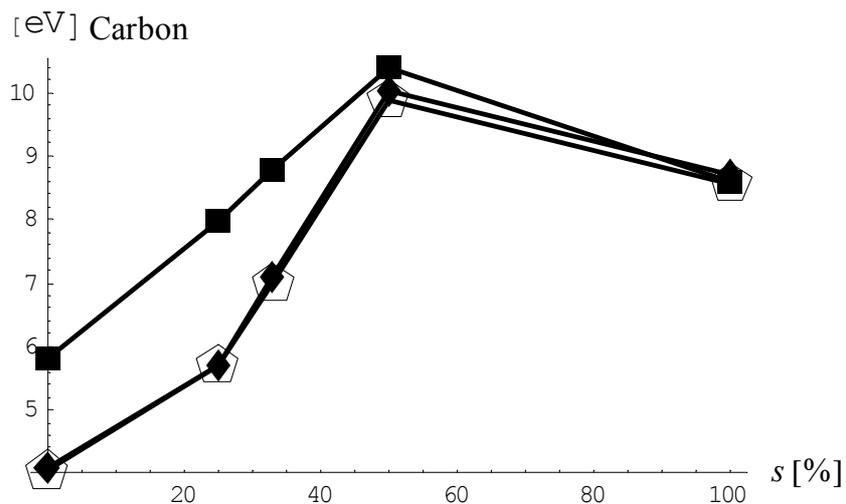
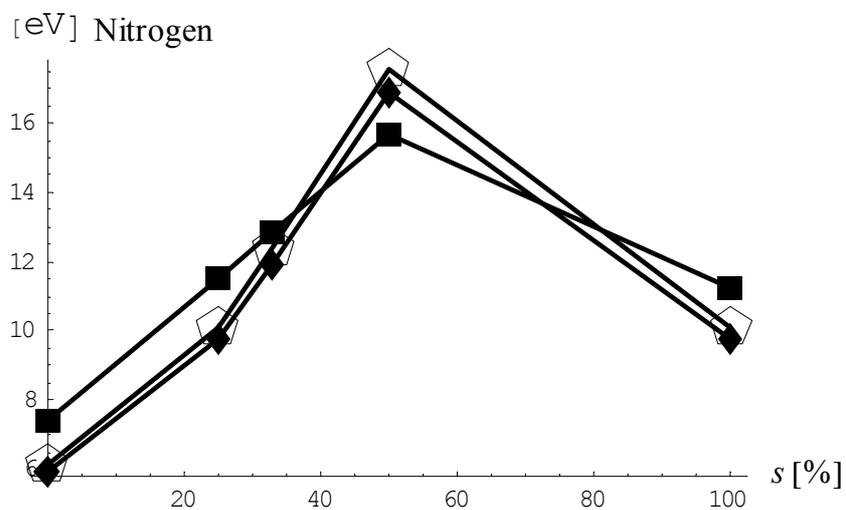
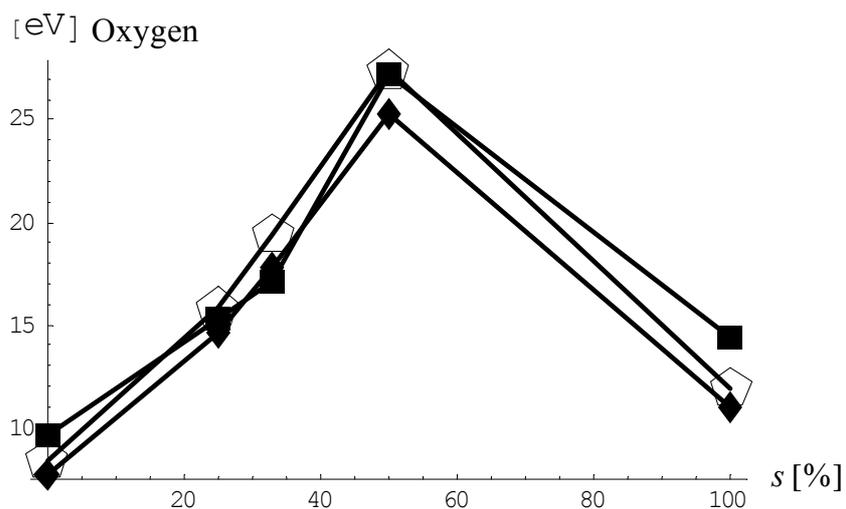

**Figure 1.** Chemical actions (♦), the Path Integral (⬠) and Mulliken-Jaffe (■) orbital electronegativities for C, N and O atoms versus the percent of s orbital.

The adopted *PI* procedure supports different model potentials. With the aim to test the reliability of the present algorithm for the Coulomb potentials the Bachelet-Hamann-Schülter pseudopotentials [59] for C, N and O atoms were adopted and their electronegativity values recalculated. The results, both for electronegativity and chemical action, are reported in Table 4.

**Table 4.** Chemical Actions ($C_A$) and Mulliken electronegativity ($\chi_M$) values for C, N and O atoms as provided by the use of Stuttgart-Bonn (SB)[a] and Bachelet-Hamann-Schülter (BHS)[b] Model Potentials.

| Potential Model | C | | N | | O | |
|---|---|---|---|---|---|---|
| | $C_A$ | $\chi_M$ | $C_A$ | $\chi_M$ | $C_A$ | $\chi_M$ |
| SB [a] | 8.69 | 8.56 | 9.73 | 10.13 | 10.93 | 11.87 |
| BHS [b] | 8.95 | 8.7 | 9.86 | 10.56 | 11.2 | 12.26 |

[a] Ref. [43].
[b] Ref. [59].

The numerical values are only slightly higher than those obtained using Bonn/Stuttgart pseudopotentials [43] and the close relation between electronegativity and chemical action values is present, in all cases.

**Conclusions**

In the framework of density functional theory, we propose the use of the Feynman-Kleinert path integral formalism in the markovian approximation to obtain an electronegativity scale. It was evidenced how this method corresponds in terms of partition function to the Levy's constrained-search formalism for the wave function. The obtained electronegativity scale follows almost all the general criteria for its acceptability.

**Appendix: Density Functional Mulliken Electronegativity Formulation**

For an *N*-electronic system placed into an external potential $V(x)$ the general (first order) equation of the change in electronegativity, $\chi = \chi[N,V(x)]$, can be written as [32]:

$$-d\chi = 2\eta dN + \int f(x) dV(x) dx \qquad (A1)$$

in which the variation of the electronegativity $\chi$ (or the negative chemical potential in the Parr definition $\mu = -\chi$ [32, 34]) for an electronic state correlates with the number of electrons and potential variation through the chemical hardness ($\eta$):

$$2\eta = -\left(\frac{\partial \chi}{\partial N}\right)_V \qquad (A2)$$

and the Fukui function (*f*), [32, 37]:

$$f(x) = -\left(\frac{\delta \chi}{\delta V(x)}\right)_N \equiv \left(\frac{\partial \rho(x)}{\partial N}\right)_V \qquad (A3)$$

being *x* the position vector.

Next, let us express the hardness and Fukui function through the relations [32, 48]:

$$2\eta = \frac{1}{S} \qquad (A4)$$

$$f(x) = \frac{s(x)}{S} \qquad (A5)$$

where *S* and *s(x)* represent the global and the local softness defined as:

$$S = -\left(\frac{\partial N}{\partial \chi}\right)_V \qquad (A6)$$

$$s(x) = -\left(\frac{\partial \rho(x)}{\partial \chi}\right)_V . \tag{A7}$$

Global and local softness are correlated among themselves through the relation,

$$S = \int_{-\infty}^{+\infty} s(x) dx \tag{A8}$$

on the base of assumed $N$-normalized density:

$$N = \int_{-\infty}^{+\infty} \rho(x) dx . \tag{A9}$$

Using the expressions A4 and A5 we can integrate the equation A1 for the electronegativity to obtain:

$$\chi(N) = -\int_0^N \frac{1}{S} dN - \frac{1}{S} \int_{-\infty}^{+\infty} s(x) V(x) dx . \tag{A10}$$

assuming the initial zero electronegativity value as $V(x) \to 0$. The integrals in A10 can be carried out once the local and global softness $s(x)$ and $S$, respectively, are analytically known. This can be achieved assuming a *quasi* independent-particle model within density functional theory providing the following expression for the softness kernel $s(x, x')$ [48]:

$$s(x, x') = -\frac{\nabla \rho(x')}{\nabla V(x')} \delta(x - x') + \rho(x)\rho(x') . \tag{A11}$$

From the softness kernel the local softness $s(x)$ can be recovered by integrating the equation A11 over $x'$:

$$s(x) = \int_{-\infty}^{+\infty} s(x, x') dx$$

$$= -\frac{\nabla \rho(x)}{\nabla V(x)} + N\rho(x) \tag{A12}$$

where the well-known delta-Dirac integration rule

$$\int_{-\infty}^{+\infty} g(x) \delta(x - x') dx' = g(x) \tag{A13}$$

and the normalization condition A9 were used.

Successively, the global softness $S$ can be analytically expressed integrating the local softness A12 over $x$, with the result:

$$S = \int_{-\infty}^{+\infty} \frac{\nabla \rho(x)}{[-\nabla V(x)]} dx + N^2 \qquad (A14)$$

where the condition A9 is taken into account.

Introducing local and global softness expressions A12 and A14 into the electronegativity, equation A10, the integrals can be analytically solved yielding the formula:

$$\chi(N) = -\frac{1}{\sqrt{a}} \arctan[\frac{N}{\sqrt{a}}] - \frac{b}{a+N^2} - NC_A \frac{1}{a+N^2} \quad . \qquad (A15)$$

In order to simplify the expression, the following definitions have been introduced:

$$a = \int_{-\infty}^{+\infty} \frac{\nabla \rho(x)}{[-\nabla V(x)]} dx$$

$$b = \int_{-\infty}^{+\infty} \frac{\nabla \rho(x)}{[-\nabla V(x)]} V(x) dx$$

$$C_A = \int_{-\infty}^{+\infty} \rho(x) V(x) dx . \qquad (A16)$$

The Mulliken electronegativity can be derived performing the average of the absolute electronegativity, equation A15, over the charge region $[N-1, N+1]$, [46], with the result:

$$\chi_M = \frac{1}{2} \int_{N-1}^{N+1} \chi(N) dN$$

$$= \frac{b+N-1}{2\sqrt{a}} \arctan\left(\frac{N-1}{\sqrt{a}}\right) - \frac{b+N+1}{2\sqrt{a}} \arctan\left(\frac{N+1}{\sqrt{a}}\right) + \frac{C_A - 1}{4} \ln\left[\frac{a+(N-1)^2}{a+(N+1)^2}\right]. \quad (A17)$$

The electronegativity expression given in the equation A17 represents our proposed working electronegativity formula, see equation 4, in the framework of density functional theory.

**Acknowledgments**

The authors gratefully acknowledge Università della Calabria and MIUR for the financial support. MVP thank Prof. Adrian Chiriac (Chemistry Department, West University of Timisoara) for helpful discussions.